\documentclass[doublecol]{epl2} 
\def\gs{\gtrsim}
\def\ls{\lesssim}
\def\be{\begin{equation}}
\def\en{\end{equation}}    
\def\ep{\epsilon}
\def\gs{\gtrsim}
\def\ls{\lesssim}
\def\ve{\varepsilon} 
\newcommand{\bi}[1]{\mbox{\boldmath$#1$}}
\newcommand{\av}[1]{\langle{#1}\rangle}
\newcommand{\AV}[1]{\bigg\langle{#1}\bigg\rangle}
\def\p{\partial}
\def\bea{\begin{eqnarray}}
\def\ena{\end{eqnarray}}

\def\gs{> \kern -12pt \lower 5pt \hbox{$\displaystyle{\sim}$}}
\def\ls{< \kern -12pt \lower 5pt \hbox{$\displaystyle{\sim}$}}

\def\hrij{\hat{\bi r}_{ij}}

\def\aQ{\stackrel{\leftrightarrow}{Q}}

\def\a1{\stackrel{\leftrightarrow}{1}}

\newcommand{\pp}[2]{\frac{\partial {#1}}{\partial {#2}}}

\title{Molecular dynamics simulation of orientational  glass 
formation in anisotropic particle systems   in three dimensions}
\shorttitle{{Simulation} of orientational glass} 
\author{Kyohei Takae and Akira  Onuki}
\shortauthor{Kyohei Takae  and Akira  Onuki}

\institute{                    
Department of Physics, Kyoto University, 
Kyoto 606-8502, Japan
}
\abstract{We propose  a simple 
microscopic model of molecular dynamics simulation 
to study orientational glass 
in three  dimensions. We present  
simulation results for    mixtures of mildly  anisotropic 
particles  and spherical impurities. 
We realize fcc  solids 
without orientational order in a rotator phase. 
As  the temperature $T$ is lowered,   
the disordered matrix is gradually replaced by   
  four kinds of orientationally ordered,  {rhombohedral} domains. 
Two-phase coexistence is realized   
in a temperature window. 
The impurities serve to 
anchor the orientations of 
the surrounding anisotropic  particles, resulting in  
finely divided  domains or medium long-range orientational order. 
We examine  the rotational dynamics of 
the molecular orientations   which is slowed  down at low $T$. 
We  predict the    shape memory effect 
  under a stretching cycle due to 
 inter-variant transformation. 
 }

\begin{document}

\maketitle

\section{Introduction}

Nonspherical  molecules 
such as KCN can form a crystal 
without  long-range orientational order for mild molecular anisotropy 
 \cite{ori}, while  liquid crystal 
phases can appear for large molecular   anisotropy. 
Such crystals  are called  plastic crystals in a rotator phase. They 
 undergo an orientational  phase transition 
 as  the temperature $T$ is further lowered, where 
  the crystal structure is 
 cubic at high $T$   and  non-cubic at low $T$.  
With inclusion  of impurities in such solids, 
 the so-called  orientational glass has been  realized \cite{ori}.  
Around the  transitions, a peak in  the specific heat \cite{ori} and  
softening of the shear modulus \cite{ori,Bell} 
have been observed.    In real systems,  the 
molecules often have dipolar moments, yielding 
dielectric anomaly.  
%where the impurities serve to pin the  heterogeneous   
%orientation fluctuations.  
 As a similar example, 
metallic ferroelectric glass, called  relaxor,  
has been studied extensively \cite{relax}, 
where  frozen polar nanodomains have been observed.  
%Recently,   a system of  
% off-stoichiometric intermetallic Ti-Ni 
%was shown to be 
%glassy martensite or strain glass,
%exhibiting the shape-memory effect 
%and the superelasticity  
%  \cite{Ren}. 

For one-component  anisotropic 
particle systems, there have been a number of simulations 
on the statics  
\cite{Frenkel1,Allen1,Singer,Vega,Radu,Jackson,Dijkstra}  
and dynamics  
\cite{Michele,new}  of the 
orientational phase transition.   
For two-component 
anisotropic particle systems, 
%nucleation from liquid to crystal 
%has been studied, 
Chong {\it et al}.  examined  the slowing-down of the 
orientational time-correlation functions 
around the glass transition \cite{Chong}. 
In this paper, setting up  a simple microscopic model, 
 we will  investigate   the formation  of 
orientational glass. 
In particular, we will examine 
 how impurities can microscopically produce   
orientational disorder, which has remained  
unexplored in the literature. Furthermore,   alignments of 
anisotropic particles formig a crystal 
 give rise to   lattice  deformations. As a unique feature in  
orientational glass, 
  heterogeneous strains 
should  emerge on mesoscopic scales.  In such situations, 
 we may predict   soft elasticity and 
a shape memory effect  against  applied stress.

\section{Model}

%In our model of molecular dynamics simulation, 
We consider a binary mixture of anisotropic  particles with number $N_1$ 
and  spherical particles  with number $N_2$. In this paper, 
we set  $N_1+N_2=8192$. 
The  composition of the second species is defined by 
  \be 
  c=N_2/(N_1+N_2). 
\en    
%which is either of 0, 0.05, 0.1, 0.15, 0.2, or 0.3 
We assume  relatively small $c$, so 
the spherical particles may be  treated as impurities. 
The particle   positions are  
written as ${\bi r}_i$ ($i=1, \cdots, N$). 
The  anisotropic  particles are assumed to be axisymmetric; then,  
their orientation  
 may be expressed in terms of the solid angles $\phi_i$ and $\theta_i$ 
($i=1, \cdots, N_1$)  as 
$
{\bi n}_i=(\sin\theta_i\sin\phi_i, \sin\theta_i\cos\phi_i, \cos\theta_i). 
$  
The particle sizes of the two species are characterized by 
lengths $\sigma_{1}$ and $\sigma_2$. 
The pair potential $U_{ij}$  between particles 
$i \in \alpha$ and $j\in \beta$ 
($\alpha,\beta=1,2$)  is a  truncated 
modified Lennard-Jones potential depending on the 
particle distance $r_{ij}= |{\bi r}_{i}-{\bi r}_j|$ 
and the directions ${\bi n}_i$ and ${\bi n}_j${.}
For  $r_{ij}> r_c=3\sigma_1$ it is  zero, while  
for  $r_{ij}< r_c=3\sigma_1$ it  reads   
\be  
U_{ij}
=4\ep\bigg[(1+ A_{ij}) 
\frac{\sigma^{12}_{\alpha\beta}}{r_{ij}^{12}}
%-(1+ B_{ij}) 
-\frac{\sigma_{\alpha\beta}^6}{r_{ij}^6} \bigg] -C_{ij}, 
\en 
where  $\ep$ is 
the characteristic interaction energy and    
\be 
\sigma_{\alpha\beta}=
(\sigma_\alpha + \sigma_\beta)/2. 
\en  
The particle anisotropy  
is taken into account by the 
 angle factor $A_{ij}$, which depends on 
 the relative direction 
$\hrij= r_{ij}^{-1}({\bi r}_{i}-{\bi r}_j)$ 
and the orientations ${\bi n}_i$ and ${\bi n}_j$. 
%There can be a variety of their forms 
%depending on the nature of the anisotropic interactions. 
In  this paper, we assume the following form,      
\be
A_{ij} = \chi \delta_{\alpha 1} 
({\bi n}_i\cdot\hrij)^2+\chi 
\delta_{\beta 1} ({\bi n}_j\cdot\hrij)^2, 
\en
where $\chi$ is the anisotropy strength of  repulsion.  The  
 $\delta_{\alpha 1}$ ($\delta_{\beta 1}$) is equal to 
1 for $\alpha=1$ $(\beta=1$) and  0 for $\alpha=2$ ($\beta=2$). 
In the right hand side, 
 the first (second) term is nonvanishing only when 
$i$ ($j$) belongs to the first species. 
The   $C_{ij}$ ensures the continuity 
of $U_{ij}$   at $r=r_c$.  
% ensuring the continuity of $U_{ij}$.  

In our system,  the total  potential energy 
is the sum  $U=\sum_{1\le i<j\le N}U_{ij} $, while 
the total kinetic 
energy $K$ arises from the translational velocity 
${\dot{\bi r}}_i= d {\bi r}_i/dt$ and the rotational 
{velocity}   ${\dot{\bi n}}_i= d {\bi n}_i/dt$ as 
\be
%U &=&  \sum_{1\le i<j\le N}U_{ij}, \\
K =  \sum_{1\le i\le N}\frac{1}{2} m_\alpha |{\dot{\bi r}}_i|^2
 +  \sum_{1\le i\le N_1} \frac{1}{2} I_1 |{\dot{\bi n}}_i|^2,
\en   
where $m_1$ and  $m_2$   are 
 the masses,   and $I_1$ is the  moment of inertia  
 of the first species. We  set $m_1=m_2$ in our simulation.  
The molecular rotation  around 
the symmetry axis parallel to ${\bi n}_i$ 
does not change $U$, so we may neglect its  kinetic energy. 
The Newton equations  of motion for translation and rotation are  written as 
\cite{Allen}  
\bea
&&{m_\alpha} 
{\ddot{\bi r}}_i =-\pp{U}{{\bi r}_i} \quad (i=1,\cdots, N), \\ 
&& I_1{\bi n}_i \times {\ddot{\bi n}}_i =  -{\bi n}_i \times 
\pp{U}{{\bi n}_i} \quad (i=1,\cdots, N_1), 
\ena  
where   
${\ddot{\bi r}}_i= {d^2}{\bi r}_i/{dt^2}$ and 
 ${\ddot{\bi n}}_i= {d^2}{\bi n}_i/{dt^2}$.  
%We solve Eq.(10) by quaternion method \cite{Allen}.
Hereafter, we measure space,   time, and temperature  
in  units of $\sigma_1$,   
$ 
\tau_0 = \sigma_1 \sqrt{m_1/\ep}, 
$ 
and $\epsilon/k_B$, respectively. 

Treating    
equilibrium or  at least 
nearly steady states,   we  furthermore 
attach  a Nos$\acute{\rm e}$-Hoover thermostat 
\cite{nose} to all the particles. That is, we 
added   the thermostat terms  
$-\zeta_{\rm NH}(t) m_\alpha {\dot{\bi r}}_i$  and 
$-\zeta_{\rm NH}(t)I_1  {\bi n}_i\times {\dot{\bi n}}_i$  in the 
right hand sides  of Eqs.(6) and (7), respectively, where 
  $\zeta_{\rm NH}(t)$ 
obeys 
\be 
\tau_{\rm NH}^2 \frac{d}{dt} {\zeta_{\rm NH}(t)}= -1+K/[k_BT(3N/2+N_1)],
\en 
 with 
$\tau_{\rm NH}$ being a short relaxation time taken to be 0.1. 
% Unless confusion may occur, 
% where $k_B$ is the Boltzmann constant. 
%Stress 
%will be measured 
%in units of $\epsilon/\sigma_1^2$. 

%Assuming that the particles of the second species 
%are  spherical and larger,    we set $\chi_2=0$,   
% $\sigma_2/\sigma_1=1.2$ or $1.4$, and   
%  $c=0, 0.1$  or $0.2$. 
%From Eqs.(2)-(4) 
 We may envisage 
the anisotropic  particles  
as spheroids depending on 
the parameter $\chi$ from Eq.(4).  For  particles $i$ and $j$ of 
the first species, 
minimization of 
$U_{ij}$  with respect to $r_{ij}$ 
 yields  $r_{ij}= 2^{1/6}(1+A_{ij})^{1/6}\sigma_1$. 
For $\chi>0$,   the  short and long  
diameters,  $a_s$  and $a_\ell$, are given by    
\be 
a_s= 2^{1/6}\sigma_{1},\quad  a_\ell=
 (1+2\chi )^{1/6}2^{1/6}\sigma_{1},  
\en 
for the perpendicular and parallel 
orientations  of  ${\bi n}_i$ and ${\bi n}_j$  
with respect to $ {\hat{\bi r}}_{ij} $, respectively. 
% the  longest  diameter 
%$a_\ell$  for the parallel orientations  
%with ${\bi n}_i\cdot {\hat{\bi r}}_{ij} 
%= {\bi n}_j \cdot {\hat{\bi r}}_{ij} =\pm 1$. Thus,   
The inertia  momentum of the first species in  Eq.(7) 
is  given by  
\be 
I_{1}= (a_\ell^2+ a_s^2)  m_1  /20. 
%\quad
%I_{1\ell}= 2a_s^2  m_1  /5.
\en  

In our simulation, we set $\sigma_2/\sigma_1=1.1$ and 
$\chi=1.2$. From Eq.(9)   the aspect ratio 
 is $a_\ell/a_s=1.23$.  For  
these mild parameter values, 
we realized   fcc plastic solids    without long-range  orientational order 
at relatively high $T$. For large $\chi$, say 10, 
we realized liquid crystal phases in our model. 
It is worth noting that our angle-dependent potential (2) is 
analogous to the Gay-Berne potential 
for anisotropic molecules used 
to simulate mesophases  of liquid crystals \cite{Gay}.  
Similar  angle-dependent potentials have also been used 
for lipids forming  membranes.  
\cite{Leibler,Noguchi}.

\section{Four-variant states  at fixed volume}

Starting with 
liquid states at $T=2$, we 
 {quenched} the system to $T=0.5$ 
below the melting  
and waited for $10^4$ to realize  fcc plastic crystals. 
%Simulation times were longer than $5\times 10^4$. 
 Here, a few 
stacking faults were formed in most runs. 
We then cooled the system to a 
final temperature and waited for $3\times 10^4$, where 
orientational order developed. In Figs.1-7, 
we fixed the system volume and shape imposing  the periodic boundary condition.
  In terms of the {molecular} volumes $v_1= 
 \pi a_s^2 a_\ell/6$  and 
 $v_2  =\pi 2^{1/2}\sigma_{2}^3/6$, the packing fraction 
$\phi_{\rm pack}$ is given by      
\be 
 \phi_{\rm pack}=(N_1v_1+ N_2 v_2)/V. 
\en    
 We set  $ \phi_{\rm pack}=0.82$ in Figs.1-7. Then the  system length  is 
 about $ 20$ and the pressure is about $4\epsilon/\sigma_1^3$. 
%The  average potentail energy 
%is about $-5 \epsilon$ per particle.   

%1
\begin{figure}
%\begin{figure}
%\onefigure{FCsig11_3.eps}
\onefigure{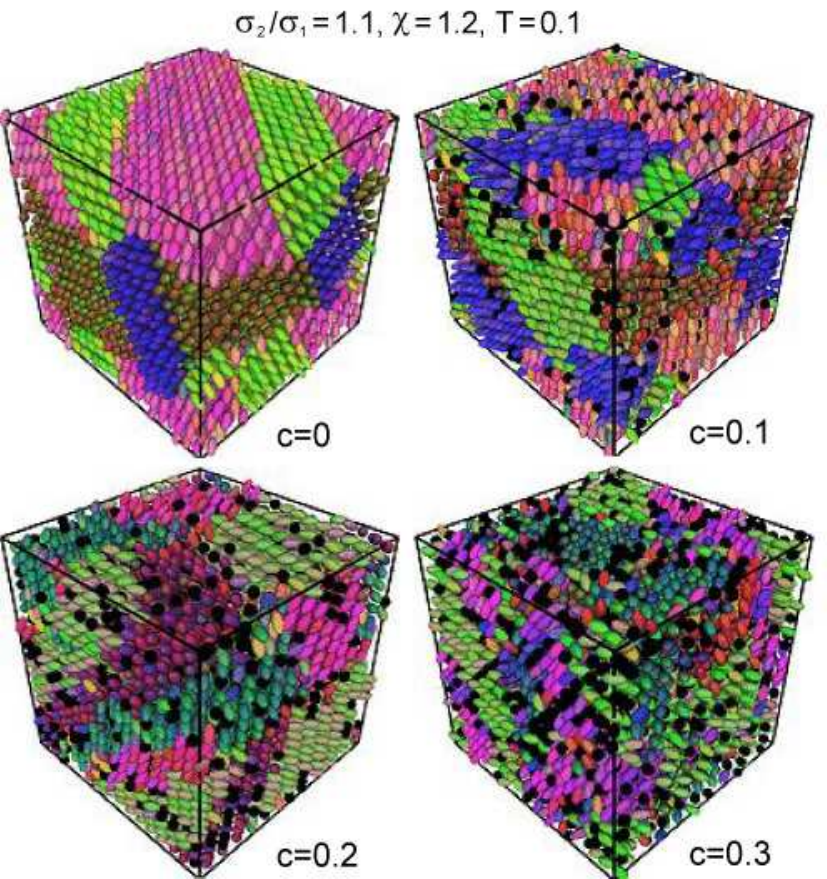}
%\onefigure{noden.eps}
\caption{Frozen domain structures 
composed of four {rhombohedral} variants 
for $\sigma_2/\sigma_1=1.1$, $\chi=1.2$, and $T=0.1$ 
at fixed volume.  With increasing the 
composition $c$ of the impurities (black points), 
the orientational disorder increases, 
leading to a decrease in 
 the domain size. 
The particle color is blue, green,  and red  for 
${\bi n}_i= (\pm 1,0,0)$, $(0,\pm 1,0)$, $(0,0,\pm 1)$, respectively.
}
\label{fig.1}
\end{figure} 
%2
\begin{figure}
%\begin{figure}
%\onefigure{anchor.eps}
\onefigure{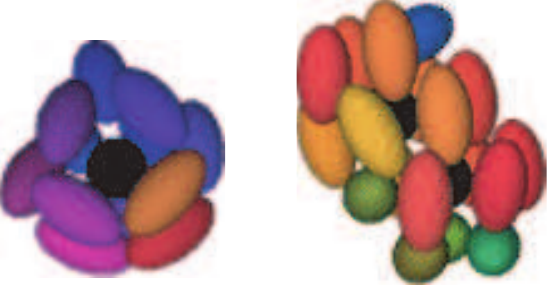}
%\onefigure{noden.eps}
\caption{Anchoring of spheroidal particles 
around a single  impurity (left) and two associated 
impurities (right) for  $\sigma_2/\sigma_1=1.1$, $\chi=1.2$, 
and $T=0.1$.}
\label{fig.1}
\end{figure} 

%3
\begin{figure}[tbp]
\begin{center}
\includegraphics[width=200pt]{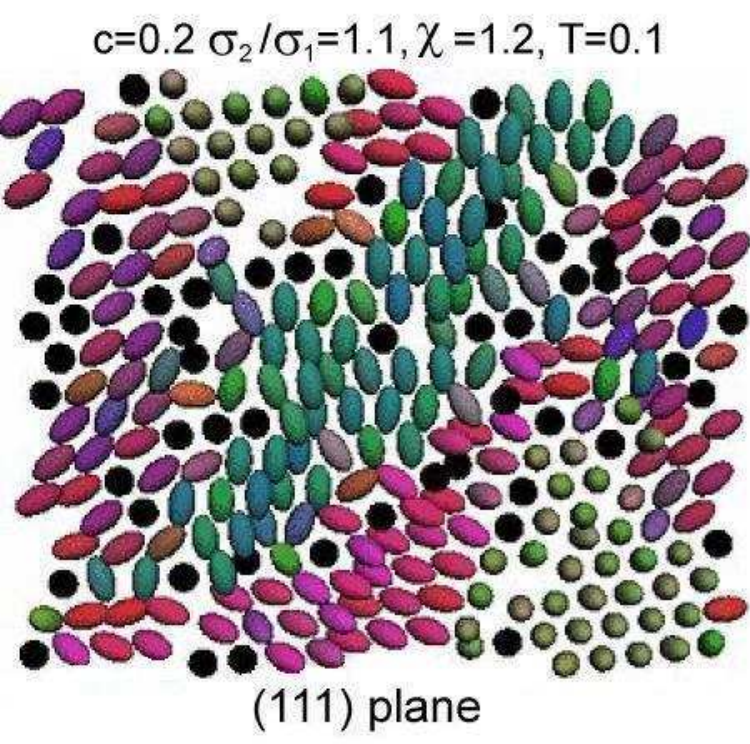}
\caption{
Crosssectional snapshot in a (111) plane, 
where four {rhombohedral} domains composed of 
spheroidal particles can be seen 
with impurities (black) disturbing the orientations. 
Here,  $c=0.2$, $\sigma_2/\sigma_1=1.1$, $\chi=1.2$, and $T=0.1$ .}
\end{center}
\end{figure}

%4
\begin{figure}
%\begin{figure}
%\onefigure{FAchi2Configuration2.eps}
%\onefigure{FImpurity_clustering.eps}
%\onefigure{FImpurity_clustering2.eps}
\onefigure{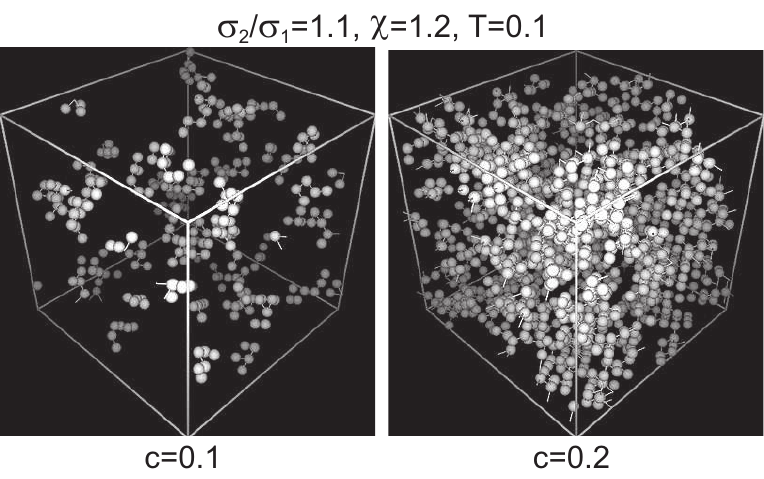}
\caption{Snapshots  
 of impurities composing clusters with 
sizes $\ge 5$ for $c=0.1$ (left) and for $c=0.2$ (right) 
with mesoscopic heterogeneities.
%Significant clustering appears. 
The data are the same as those in Fig.1. 
 }
\label{fig.1}
\end{figure}

In  Fig.1, the orientational  domains can be seen at  $T=0.1$, where  
 the particle positions and orientations are nearly frozen in time. 
For  $c=0$,  the system undergoes a structural 
phase {transition}, where  
the lattice structure changes from {fcc} to 
rhombohedral almost without dilation  change. 
The lattice constant is equal to $a=1.65$  through the transition. 
The ordered phase consists of  four rhombohedral variants 
where the angles of 
the rhombuses are $83^\circ$ and $97^\circ$.
% and the side length remains equal to $a=1.65\sigma_1$. 
The separation distance of the closely packed 
 $(111)$ planes is increased   from $a/\sqrt{3}=0.952$ to  1.08,  because 
the orientation vectors ${\bi n}_i$ are aligned  in the $[111]$ direction.  
With increasing $c$,  
the orientational disorder is gradually 
increased, leading to pinning of finely divided domains. 
We  eventually obtain a glass state at $c=0.3$.  

In  Fig.2,  
the anisotropic 
particles    are  
aligned in the perpendicular directions $(\perp {\hat{\bi r}}_{ij}$) 
around  one or two  impurities. This parallel anchoring 
disturbs the  orientation order.  
In   Fig.3,    in a (111) plane, we  display   
the  four {rhombohedral} variants with different orientations  for $c=0.2$.
The interfaces between the variants tend to be trapped at clustered  
 impurities.    
We also notice that the junction angles, at which two or more 
domain boundaries intersect,
are approximately multiples of $\pi/6$. Similar   patterns were   observed 
on hexagonal planes in a number of 
experiments on alloys after  
structural phase transitions \cite{Kitano}.

\section{Impurity clustering}   

Figure 4 shows   significant impurity 
clustering, which  appeared 
 during quenching from liquid. 
The average impurity number per  cluster  
is   5.5 for $c=0.1$ and there is  a big 
percolated cluster for $c=0.2$. Here, 
 two impurities are treated to belong to the same cluster 
 if  their  distance is  
smaller than 1.3. In the snapshot of $c=0.3$ in Fig.1, 
the impurity clustering 
is closely related to  
 the   heterogeneity 
in  the orientations.

To explain this effect, we compare  the solvation energy 
of a single impurity  and that of two  associated impurities  
(see  Fig.2).  In terms of 
  the potential energy $U_j= \sum_k U_{jk}/2$ of particle $j$, 
it  is estimated as 
\be 
U_{\rm sol} =  U_{\rm imp}+ \sum_{{\rm nearby}~j\in 1}(U_{j}-\bar{U}_1).
\en  
Here,   $ U_{\rm imp}$ is the contribution from  
the impurities  under consideration, 
 the summation is over the nearby anisotropic 
 particles $j$ with  $r_{ij}<1.3$ ($i\in 2$  and 
$j \in 1$),  and 
$\bar{U}_1$ is the average potential energy  of the 
non-neighbor  anisotropic particles 
separated from any impurities longer  than $1.3$. 
%Here $\bar{U}_1$ is about $-4.5\epsilon$ at $T=0.1$ 
%and is about $-3.9\epsilon$ at $T-0.25$.  
At   $T=0.1$, 
% and $c=0.1$, 
$U_{\rm sol}$  is calculated as $U_{\rm sol}^{(1)}=-6.3\epsilon$ 
  for a single  impurity 
and $U_{\rm sol}^{(2)}=-13.8\epsilon $  
for a dimer, where   $\bar{U}_1\cong -4.5\epsilon$. The difference  
$\Delta U_{\rm sol}= U_{\rm sol}^{(2)}-2U_{\rm sol}^{(1)} 
(\sim -1.2 \epsilon)$ 
is  the association energy.  At    $T=0.25$, 
we have   $\Delta U_{\rm sol}\sim -1.8\epsilon$.  
% only weakly 
%depends on $T$ below the melting, indicating a tendency of 
%clustering in liquid. 
The total potential energy is  thus  
 lowered   with impurity clustering  in the present case. 

%Let us treat two impurities  belong to  the same cluster 
%if there distance is shorter than $1.3\sigma_1$. 
%Let $M_\ell$ be the cluster number with size $\ell$ 
%with $\sum_\ell \ell M_\ell= N_2$. 
% For $c=0.1$,  the average impurity number per  cluster  
%$\av{\ell}= \sum_\ell \ell^2 M_\ell/ N_2$ 
%is   5.5.   For $c=0.2$,  it is about 700  because of the presence of a 
%percolated  impurity cluster 
%composed of about 900 particles. 
%The relation between the compositional heterogeneities 
%and the nanodomains 
%have been studied for relaxor ferroelectrics \cite{Cluster}. 

% \sum P_^ell= cluster number
% \sum \ell P_\ell =cN
% \sum \ell^2 /\sum P_ell = ,<\ell^2>
  
%5
\begin{figure}[tbp]
\begin{center}
\includegraphics[width=200pt]{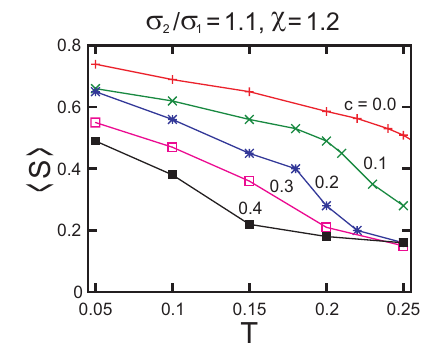}
\caption{$\av{S}=\sum_{i\in 1} S_i/N_1 $ vs $T$,  
for $c=0,0.1,0.2,0.3$, and 0.4 
for  $\sigma_2/\sigma_1=1.1$, {and $\chi=1.2$.} 
This quantity represents  the overall degree of orientational order.  
}
\end{center}
\end{figure}

%6
\begin{figure}[tbp]
\begin{center}
\includegraphics[width=241pt]{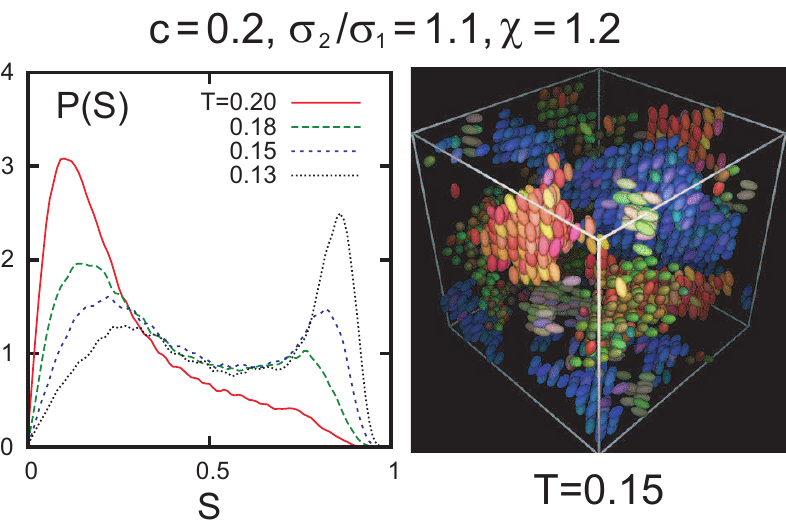}
\caption{Left: Distribution 
of the orientation amplitude $P(S)$  
in Eq.(15)  for various $T$, where 
 $c=0.2$,   $\sigma_2/\sigma_1=1.1$, and $\chi=1.2$. 
It exhibits  double peals for $T=0.13$, 0.15, 
and 0.18  representing coexistence of 
cubic and {rhombohedral} regions. Right: Snapshot of  
 anisotropic particles with $S_i >0.6$ at  $T=0.15$,   
forming  ordered domains embedded in  
a disordered matrix.   }
\end{center}
\end{figure}

\section{Coarse-grained 
orientation order parameter}

For each particle $i$ 
of the first species,  we    introduce the orientation tensor ${\aQ}_i =
 \{ Q_{i\mu\nu}\}$ ($\mu,\nu=x,y,z$) as 
%in terms of the orientation vectors ${\bi n}_k$  as 
\be
\aQ_i=
\frac{1}{1+n_{\rm b}^i} \bigg({\bi n}_i{\bi n}_i
 + \sum_{j\in {\rm bonded}} {\bi n}_j {\bi  n}_j\bigg) 
-\frac{1}{3}\a1,  
\en 
where ${\a1}= \{\delta_{\mu\nu}\}$ is the unit tensor.  
%${\bi d}_i$ is  the director with $|{\bi d}_i|=1$.  
The summation is 
 over the bonded particles  of the first species  
with $|{\bi r}_{ij}| <1.5\sigma_1$ and  
 $n_{\rm b}^i$ is  the number of 
these   particles. If a  fcc  lattice 
is formed, it   includes  
 the   nearest neighbor particles, so $n_{\rm b}^i \sim 12$.  
%The amplitude $q_i$ is  given by 
%\be 
%q_i^2= 2 \sum_{\mu,\nu}  Q_{i\mu\nu}^2/N_1.
%\en 
We  define the amplitude of the orientational order for each anisotropic 
particle $i$ as   
\be 
{S_i}={\frac{3}{2}} \sum_{\mu,\nu}  Q_{i\mu\nu}^2. 
\en   
Here,   $S_i\sim 0.1$  in disordered regions 
 due to the thermal fluctuations,  
but it increases  up to unity within {rhombohedral} domains 
at low $T$. 
In Fig.5,  the  average 
$\av{S}=\sum_{1\le i \le N_1} S_i/N_1$ 
 represents the overall degree of orientation order.  
In our simulation, we  realized  only uniaxial  states, where 
we have $\aQ_i=S_i^{1/2} ({\bi d}_i {\bi  d}_i -\a1/3)$ in terms of 
the amplitude  $S_i$  and the director  
${\bi d}_i$.

\section{Coexistence of high and low temperature phases} 

In Fig.5, the degree of orientational 
order $\av{S}$ increases    continuously as $T$ is lowered. 
For small $c$ at fixed volume and shape,  however, 
we find coexistence of cubic and rhombohedral  regions 
 in a {temperature} window, 
which is {roughly} given by  $0.26 \ls T\ls 0.29$ for $c=0$ 
and $0.14 \ls T\ls 0.19$ for $c=0.2$.    
To examine this  coexistence, 
we introduce the distribution of the orientational amplitude, 
\be 
P(S)=\frac{1}{N_1} \sum_{1\le i\le N_1} \av{\delta (S-S_i)}.   
\en 
Taking the average  $\av{\cdots}$   over six  runs 
we obtained  $P(S)$ in Fig.6, which  exhibits    two peaks 
 for $0.13\le T\le 0.18$ at $c=0.2$. 
We  also give  a snapshot of the ordered anisotropic particles with $S_i>0.6$. 
For $c=0$,   the interfaces between the ordered and disordered regions 
are temporally  fluctuating with small amplitudes. 
Small amounts of 
impurities can  pin  the interfaces  
and  the ordered regions 
are stabilized  far from the impurity clusters.

\section{Rotational dynamics}

In crystal and glass,  elongated  particles  often  undergo   
 turnover motions,    ${\bi n}_i \to  
-{\bi n}_i$. These events take   place 
in a  microscopic time $(\sim 1$), while   
the characteristic time $\tau_1$ between successive turnover motions 
 grows at low $T$. Let  us  consider  the following 
angle relaxation  function, 
\be 
W(\zeta,t) = \frac{1}{N_1} \sum_{1\le i \le N_1}  
\AV{\delta(\zeta-{\bi n}_i(t+t_0)\cdot{\bi n}_i(t_0))},
\en 
where  the average $\av{\cdots}$ is taken over 
the initial time $t_0$ and over six  runs in this paper. 
Furthermore, in terms of the $\ell$-th order Legendre polynomials 
 $P_\ell$,  we may define 
the  $\ell$-th order rotational relaxation functions as
\cite{Michele,new,Chong}   
\be
C_\ell(t)=\int_{-1}^{1}  d\zeta P_\ell(\zeta )W(\zeta,t)  . 
\en
%Here,  $C_\ell(t, \varphi)$ tends to 
%$\delta(\varphi)$ as $t \to 0$ 
%and is  broadened for $t>0$. 
Here $C_1(t)$ decays on the timescale of   $\tau_1$, while 
 $C_2(t)$ is unchanged 
by the  turnover motions. 
In orientational glass, the  ultimate 
   relaxation  of $C_2(t)$ is     
 due to the orientational 
 configuration changes involving the 
  surrounding particles, so its relaxation time $\tau_2$ 
much exceeds $\tau_1$.

%7
\begin{figure}[htbp]
\begin{center}
\includegraphics[width=240pt]{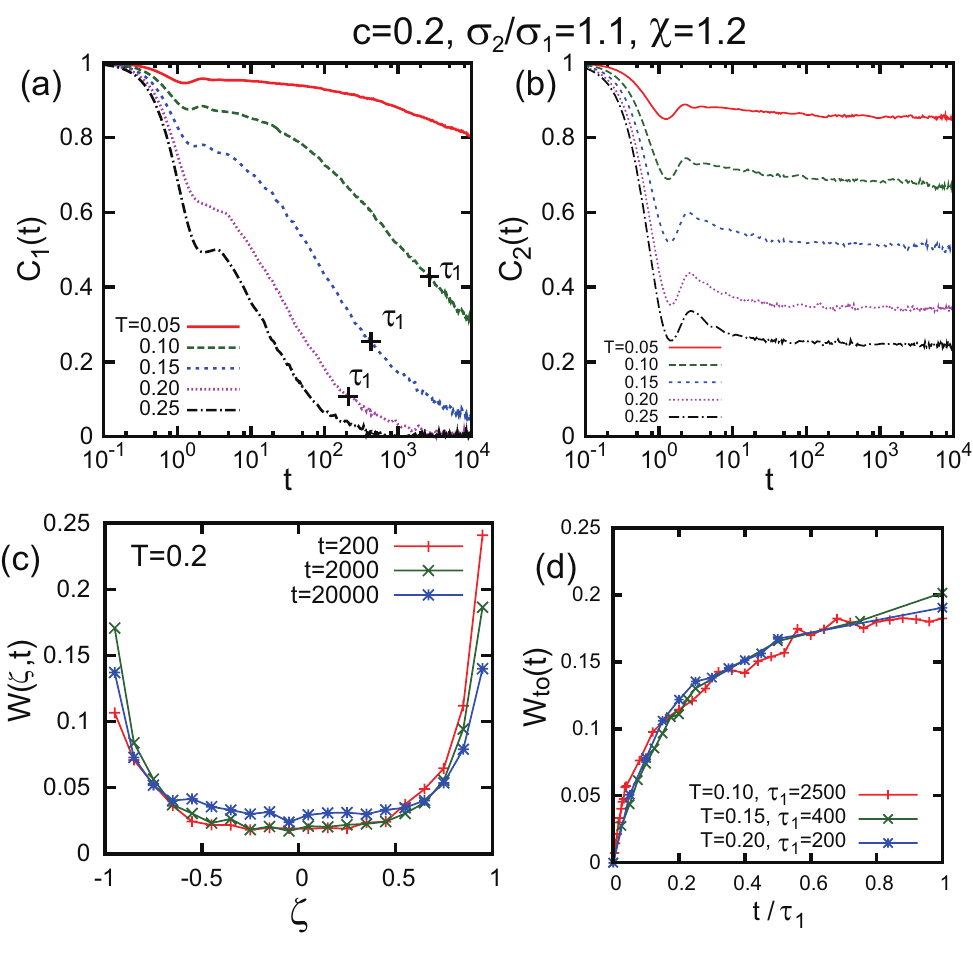}
\caption{ Orientation relaxation functions 
$C_1(t)$ in (a) and $C_2(t)$ in (b) 
  for $c=0.2$, $\sigma_2/\sigma_1=1.1$, 
and $\chi=1.2$.  $C_1(t)$ decays due to 
turnover motions and its relaxation time $\tau_1$ 
is marked on each curve (see Eq.(19)).   $C_2(t)$ can 
relax only due to the orientational configuration changes. 
(c) Distribution $W(\zeta,t)$ in Eq.(16) 
 and (d) turnover fraction  in Eq,(18)  at three times. 
%$t= 200, 2000$, and 20000. 
 }
\end{center}
\end{figure}

In  Fig.7,  
we plot   $C_1(t)$  and  $C_2(t)$ 
vs $t$ for $c=0.2$. For $t\ls 1$ they relax 
considerably due to the thermal  motions. 
The decay of $C_1(t) $ is slower for 
lower $T$, while   $C_2(t)$ apparently 
 tends to a  positive  
 constant $f_2$ for  large $t$, as in the previous 
simulation  \cite{Chong}. In our case, this plateau appears 
 because the anchoring 
of the anisotropic  particles 
around the impurities  becomes  nearly permanent 
at low $T$. Thus 
$f_2$ increases with lowering $T$.  
In  Fig.7,  we  also  present  the time-evolution of 
the relaxation function    
$W(\zeta,t)$ for $t=200, 2000$, and 20000, which  
 is peaked at $\zeta=\pm 1$. 
The peak height at $\zeta=-1$ increases  in time from 0. 
Also shown is the turnover probability  defined by 
\be 
W_{\rm to}(t)= 
\int_{-1}^{-1+ \Delta\zeta}  d\zeta~ W(\zeta,t), 
\en    
 where  we set $\Delta\zeta=0.2$. 
We notice that $W_{\rm to}(t)$ grows linearly in time 
in the very early  stage. Thus we  may define $\tau_1=\tau_1(T)$ 
from the following  linear form,  
\be 
W_{\rm to}(t)\cong \tau_1^{-1}t, 
\en   
which holds for $t/\tau_1 \ls 0.1$.  
For  $0.1 \ls t/\tau_1 \ls 1$,  
the probability of multiple turnovers becomes appreciable 
resulting in a deviation from  the linear behavior {(19)}. 
However, the scaling form $W_{\rm to}(t)= f_{\rm to}(t/\tau_1)$ 
fairly holds  for $t/\tau_1 < 1$. 
For  $t/\tau_1 \gs 4$, the orientational structural relaxation 
 comes into play 
and the scaling in terms of $\tau_1$  does not hold.

\section{Shape memory effect}

In the presence of  multi-variant orientational order,     the 
  shape-memory effect (mechanical hysteresis) emerges almost without 
dislocation formation. This effect 
 is  well-known for  shape memory alloys such as Ti-Ni 
\cite{Ren}, where 
a structural phase transition is caused by 
 the relative atomic displacements 
 in each unit cell.  
In our case,  under stretching along the $z$ axis,    
domains with ${\bi n}_i$ nearly 
parallel to the $z$ axis 
 grow yielding an  increase in  the strain. 
   We are not aware of any 
experiments on this effect in 
  anisotropic particle systems. 
%8
\begin{figure}[htbp]
\begin{center}
\includegraphics[width=250pt]{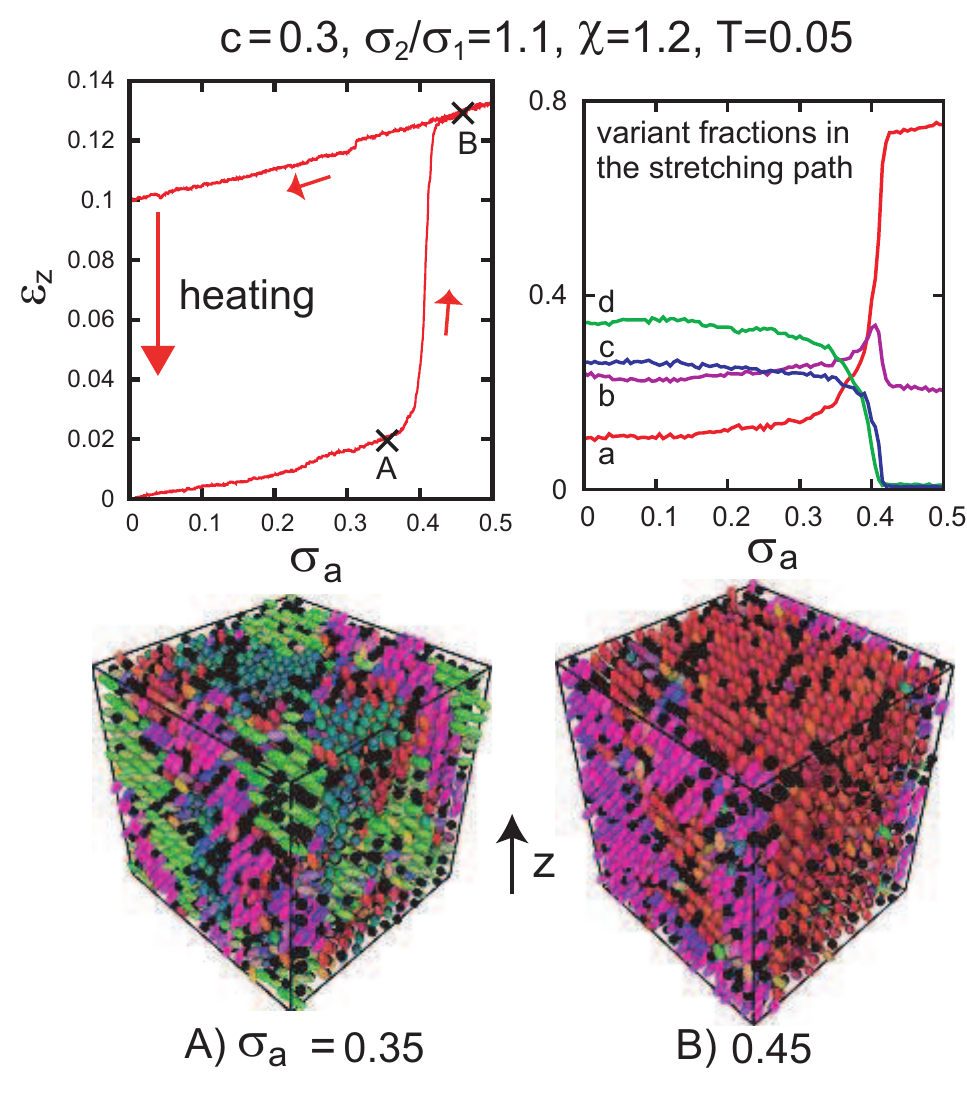}
\caption{Shape memory effect 
under  stretching  along the $z$ axis in orientational glass  
with $c=0.3$ in stress-controlled simulation, where  $T=0.05$, 
$\sigma_2/\sigma_1=1.1$,  and $\chi=1.2$. 
Top left: Strain $\ve_z$ vs  stress ${\sigma_{a}}$, 
where   
$\varepsilon_z$  increases slowly 
in the hard  ranges  $0<\sigma_a<0.38$ and $0.42<\sigma_a<0.5$    
and steeply in the soft range  $0.38<\sigma_a<0.42$. 
The return path is reversible with large Young's modulus. 
 After this cycle, the residual strain is $0.1$, 
which vanishes  upon heating to  $T=0.2$.  
Top right: 
Time-evolution of   four variant fractions on the stretching path. 
For $\sigma_a>0.38$ the two variants 
  with their  orientations ${\bi n}_i$   
nearly parallel to  the $z$ axis become dominant.
Bottom: Snapshots of the particle configurations
at (A) $\sigma_a=0.35$  and (B) $\sigma_a=0.45$.  
}
\end{center}
\end{figure}

 In   this mechanical  effect, the system shape changes. 
Thus we  assumed 
  a Parrinello-Rahman barostat \cite{Allen,Rahman} as well as 
the  Nos$\acute{\rm e}$-Hoover thermostat.  
%in  the periodic boundary condition. 
Before streching we prepared a multi-variant initial state at $t=0$,  
where the packing fraction $\phi_{\rm pack}$ in Eq.(11) was  
0.75 and the pressure was  zero. For $t>0$ we  applied the  stress    
$
 \sigma_a\equiv 
-\av{\Pi_{zz}}, 
$  
 setting   $\av{\Pi_{xx}}= 
\av{\Pi_{yy}}=0$,  where  $\Pi_{\mu\nu}$ 
are the stress components  with  
 $\av{\cdots}$ being  the space average. 
We allowed the system to take a rectangular shape.  
In terms of the  system length  $L_z(t)$  
 along the $z$ axis, the average strain is   
\be 
\varepsilon_z= L_z(t) /L_{z}(0) -1 .
\en 
%where $L_{z0}$ is   at $t=0 $ andis that for $t>0$. 
We define effective 
Young's  modulus  $E_e$  by 
$\p\varepsilon_z/\p \sigma_a=1/E_e$. 
It follows the effective shear modulus $\mu_e$ from  the formula 
$E_e= 3\mu_e/(1+\mu_e/3K)$ 
 in  classical elasticity, where 
$K$ is the bulk modulus. 
In our case $K$ is about $50$
%$33$ 
and is  much larger than $\mu_e$, 
so $E_e\cong 3\mu_e$. 
Hereafter,   $\sigma_a$ 
and $E_e$  will be   measured in units of 
$\epsilon/\sigma_1^3$.

In Fig.8, we first increased $\sigma_a$  at 
$d\sigma_a/dt=5\times 10^{-5}$ 
from 0 up to 0.5  at $T=0.05$. 
In the left panel,   we find  $E_e\sim 17$ 
for $0<\sigma_a<0.38$,  $E_e \sim 3$    for 
 $0,38 <\sigma_a<0.42$, and $E_e\sim 20$ for $0,42 <\sigma_a<0.5$.  
In the second  range,  the solid is soft 
because  the variants elongated along the $z$ axis grow     
due to inter-variant transformation. 
Here, the orientation vectors 
${\bi n}_i=(n_{ix}, n_{iy}, n_{iz})$ of the four variants  are 
{roughly} given  by 
(a) (0.3,-0.1,0.9), (b) (-0.7,0.1,0.7), (c) (0.7,0.7,0.2), and 
(d) (0.5,-0.9,0.1) in Fig.8. 
In  the right panel,   
the variant (a) grows, the variant  (b) 
remains nonvanishing, and the variants 
(c) and (d) disappear.  
Secondly, we  decreased $\sigma_a$   back to 
0 at $d\sigma_a/dt=-5\times 10^{-5}$. 
In this return path,  
$E_e$ was  about $20$,  the disfavored variants no more appeared,  
 and  a remnant strain of order  
$0.1$ remained. 
Thirdly, at $\sigma_a=0$, 
we increased $T$ from 0.05 to 0.2 into  the disordered phase. 
After  this heating,   
the system became isotropic. 
% In contrast, 
%  in the one-component case  in Fig.6, 
% we have found  no such changes 
% once a single-variant state is realized.  

\section{Summary} 

We have siudied  the orientational glass 
using the 
potential (2). Due to the anisotropic 
factor $A_{ij}$, 
 the anisotropic particles have the aspect ratio $(1+ 2\chi)^{1/6}$. 
 In our simulation,  the anisotropy parameter $\chi$ is 1.2 and 
the size ratio $\sigma_2/\sigma_1$  is 1.1. As a result,   
 fcc plastic crystals appear at relatively high $T$, 
which undergo the orientation transition at lower $T$. 
The  impurities serve  to 
pin  the orientations of the surrounding 
anisotropic particles.

We briefly summarize our main results. 
In Fig.1, we have illustrated    four-variant 
ordered states for various  $c$ at $T=0.1$. 
With {increasing} $c$,  {rhombohedral} domains  
are finely divided and their typical size is {decreased}.  
In Fig.4, we have shown that the impurities are 
heterogeneously distributed. 
In Fig.5, we have plotted  the average orientational amplitude 
$\av{S}$, which increases 
continuously with lowering $T$.
In Fig.6, we have shown   large-scale    coexistence of the 
disordered and  ordered phases  
in a temperature window. 
 In Fig.7, 
the  orientational time-correlation functions 
have been plotted for $c=0.2$, where $C_1(t)$ decays due to turnover motions. 
In Fig.8, we have examined 
the shape memory effect  in 
orientational glass.  

We next make some remarks below.\\ 
 (i)  
For small $c$,  we need to know how 
the characteristic domain size is determined. 
For moderate impurity concentrations ($c \gs 0.2$ in this work), 
the orientational disorder is proliferated. 
These aspects  should be further studied.  

\noindent (ii) 
The phase transition  delicately depends on whether 
simulations are performed at fixed volume or at fixed  stress. 
Some salient features  at fixed stress are as follows. 
(a) For $c=0$, a  first-order phase transition 
occurs with a shape change.    
 (b) For $c>0$,  both  multi-domain states 
and single-domain states can be 
realized for the same parameters as in Fig.8. 
(c) For $c>0$, the two-phase 
 coexistence in Fig.6 can still be realized even at  
fixed stress. 
%Without impurities ($c=0$),  it emerged   
%at  fixed volume,  but it eventually disappeared   
%at  fixed stress. Thus, at $c=0$, 

\noindent (iii)  
 In our simulation, the system keeps  the crystalline  order. 
However, for larger size ratio $\sigma_2/\sigma_1$, say 1.4, we found 
an increase in the positional disorder leading to 
polycrystal and positional glass.  
Competition of positional and orientational 
glass transitions 
can then be studied.  
 For large anisotropic parameter $\chi$, 
we may also study complicated phase  behavior of 
two-component liquid crystals.   
%Phase ordering  in two-component 
%liquid crystal systems  is  also worth studying. 

\noindent (iv) In this paper, the impurities 
are  spherical and slightly larger than the anisotropic particles. 
By modifying the potential form, 
we may also treat other types of impurities. 
For example, their attractive 
interaction with the host 
anisotropic particles may be anisotropic; then, 
 the anchoring can be homeotropic.

%We may furthermore   include the dipolar 
%{interaction} in our model to examine 
%the so-called relaxor behavior \cite{relax}. 
%The response to  electric field is also anomalously strong 
%in  multi-variant states of  anisotropic particle systems. 

\acknowledgments
This work was supported by Grant-in-Aid 
for Scientific Research  from the Ministry of Education, 
Culture,  Sports, Science and Technology of Japan. 
K.T. was supported by the
Japan Society for Promotion of Science.
The numerical calculations were carried out on SR16000 at YITP in Kyoto
-University.


\begin{thebibliography}{0}


\bibitem{ori} 
\Name{H\"{o}chli U. T., Knorr K. \and Loidl A.}
\REVIEW{Adv. Phys.}{39}{1990}{405}.

%\bibitem{Yamamuro} O. Yamamuro, H. Yamasaki, Y. Madokoro,
%I. Tsukushi,  and T.  Matsuo,  J. Phys.: Condens. Matter 
% {\bf 15}, 5439 (2003). 



%\bibitem{Sherwood} 
%{\it 
%The Plastically crystalline state: 
%orientationally disordered crystals},   
%edited by John N. Sherwood  
%(John  Wiley $\&$ Sons, Chichester, 1979). 
%In Chap.2, some singular results of creep experiments 
%are given around the orientation transition. 
%\bibitem{specific} 
%\Name{Mertz B. \and Loidl A.}
%\REVIEW{Europhys. Lett.}{4}{1987}{583}.


\bibitem{Bell}
\Name{Lynden-Bell R. M. \and Michel K. H.}
\REVIEW{Rev, Mod. Phys.}{66}{1994}{721}.

%\bibitem{sound1} 
%\Name{Knorr K., Volkmann U. G. \and Loidl A.}
%\REVIEW{Phys. Rev. Lett.}{57}{1986}{2544}; 
%Here the shear modulus 
%is measured in (KBr)$_{1-x}$(KCN)$_x$.
%\bibitem{sound}
%\Name{Fossum J. O. \and Garland C. W.}
%\REVIEW{J. Chem. Phys.}{89}{1988}{7441}. 


\bibitem{relax} 
\Name{Vugmeister B. E. \and Glinchuk M. D.}
\REVIEW{Rev. Mod. Phys.}{62}{1990}{993}; 
%\bibitem{Hirota}
\Name{Hirota K., Wakimoto S. \and Cox D. E.}
\REVIEW{J. Phys. Soc. Jpn.}{75}{2006}{111006}. 
 

%phase diagram 
% 
\bibitem{Frenkel1} 
%\Name{Frenkel D., Mulder B. M. \and McTague J. P.}
%\REVIEW{Phys. Rev. Lett.}{52}{1984}{287};
\Name{Frenkel D. \and Mulder B. M.}
\REVIEW{Mol. Phys.}{55}{1985}{1171};
\Name{Veerman J. A. C. \and Frenkel D.}
\REVIEW{Phys. Rev. A}{41}{1990}{3237};
\Name{Bolhuis P. \and Frenkel D.}
\REVIEW{J. Chem. Phys.}{106}{1997}{666}.  

\bibitem{Allen1} 
\Name{Allen M. P. \and Imbierski A. A.}
\REVIEW{Mol. Phys.}{60}{1987}{453}.

\bibitem{Singer}
\Name{Singer S. J. \and Mumaugh R.}
\REVIEW{J. Chem. Phys.}{93}{1990}{1278}.

\bibitem{Vega}
\Name{Vega C., Paras E. P. A. \and Monson P. A.}
\REVIEW{J. Chem. Phys.}{96}{1992}{9060};
%\Name{Vega C., Paras E. P. A. \and Monson P. A.}
%\REVIEW{J. Chem. Phys.}{97}{1992}{8543};
\SAME{97}{1992}{8543}. 
%\bibitem{Rolf} 
%\Name{Vega C. \and Monson P. A.}
%\REVIEW{J. Chem. Phys.}{107}{1997}{2696}.

\bibitem{Radu} 
\Name{Radu M., Pfleiderer P. \and Schilling T.}
\REVIEW{J. Chem. Phys.}{131}{2009}{164513}. 

\bibitem{Jackson}
\Name{McGrother S. C., Williamson D. C. \and Jackson G.}
\REVIEW{J. Chem. Phys.}{104}{1996}{6755}.
\bibitem{Dijkstra} 
\Name{Marechal M. \and Dijkstra M.}
\REVIEW{Phys. Rev. E}{77}{2008}{061405}.  


\bibitem{Michele}
\Name{De Michele C., Schilling R. \and Sciortino F.}
\REVIEW{Phys. Rev. Lett.}{98}{2007}{265702}.  



\bibitem{new} 
\Name{Caballero N. B., Zuriaga M., Carignano M. \and Serra P.}
\REVIEW{J. Chem. Phys.}{136}{2012}{094515}.

%Dynamics 
\bibitem{Chong} 
\Name{Chong S.-H., Moreno A. J., Sciortino F. \and Kob W.}
\REVIEW{Phys. Rev. Lett.}{94}{2005}{215701};  
%\bibitem{Chong1} 
\Name{Chong S.-H. \and Kob W.}
\REVIEW{Phys. Rev. Lett.}{102}{2009}{025702}. 

%\bibitem{color} color



\bibitem{Allen}
\Name{Allen M. P. \and Tildesley D. J.}
\Book{Computer Simulation of Liquids}
\Publ{Clarendon Press, Oxford}
\Year{1987}.
\bibitem{nose}
\Name{Nos\'e S.}
\REVIEW{Mol. Phys.}{52}{1984}{255}.



\bibitem{Gay}
\Name{Gay J. G. \and Berne B. J.}
\REVIEW{J. Chem. Phys.}{74}{1981}{3316};
\Name{Brown J. T., Allen M. P., del Rio E. M. \and de Miguel E.}
\REVIEW{Phys. Rev. E}{57}{1998}{6685}.
%M.P. Neal and  A.J. Parker, 
%Chem. Phys. Lett. {\bf 294}, 277 (1998); 
%R. Memmer, Liquid Crystals {\bf 27}, 533 
%(2000). 




\bibitem{Leibler} 
\Name{Drouffe J.-M., Maggs A. C. \and Leibler S.}
\REVIEW{Science}{254}{1991}{1353}. 
\bibitem{Noguchi}
\Name{Noguchi H.}
\REVIEW{J. Chem. Phys.}{134}{2011}{055101}.



\bibitem{Kitano} 
%\bibitem{Sinclair} 
%R. Sinclair and J. Dutkiewicz, 
%Acta Metell. {\bf 25}, 235  (1977);  
\Name{Kitano Y., Kifune K. \and Komura Y.}
\REVIEW{J. Phys. (Paris)}{49}{1988}{C5-201}; 
 %K. Muraleedharan, D. Banerjee, S. 
%Banerjee and S. Lele, 
% Phil.  Mag.  A, {\bf 71}, 1011 (1995); 
%\bibitem{Mano} 
\Name{Manolikas C. \and Amelinckx S.}
\REVIEW{Phys. Stat. Sol. (a)}{60}{1980}{607}. 
%\SAME{61}{1980}{179}.

%\bibitem{Chen}
%\Name{Wen Y. H., Wang Y. \and Chen L. Q.}
%\REVIEW{Phil. Mag. A}{80}{2000}{1967}.    
%Y. H. Wen, Y. Wang, L. A. Bendersky, and L. Q. Chen, 
%Acta Mater. {\bf 48}, 4125  (2000). 


%nucleation 

%\bibitem{Cluster} 
%\Name{Jin H. Z., Zhu J., Miao S., Zhang X. W. \and Cheng Z. Y.}
%\REVIEW{J. Appl. Phys.}{89}{2001}{5048}; 
%\Name{Burton B. P., Cockayne E. \and Waghmare U. V.}
%\REVIEW{Phys. Rev. B}{72}{2005}{064113}. 



\bibitem{Ren}
\Name{Sarkar S., Ren X. \and Otsuka K.}
\REVIEW{Phys. Rev. Lett.}{95}{2005}{205702};
\Name{Wang Y., Ren X. \and Otsuka K.}
\REVIEW{Phys. Rev. Lett.}{97}{2006}{225703}.
%shape memory

%; L.E. Cross, Ferroelectrics {\bf 76} 241 (1987).
%\bibitem{CowleyReview}
%R. A. Cowley,
% S. N. Gvasaliya, S. G. Lushnikov, 
% B. Roessli,  and G. M. Rotaru, Adv. Phys. 
% {\bf  60}, 229 (2011). 



%%%%
\bibitem{Rahman} 
\Name{Parrinello M. \and Rahman A.}
\REVIEW{J. Appl. Phys.}{52}{1981}{7182}.


%\bibitem{Cowley} R. A. Cowley, Phys. Rev. B {\bf 13}, 4877 (1976).  




%\bibitem{Bond}
%S.D. Bond, B.J. Leimkuhler, and B.B. Laird, J. Comput. Phys. {\bf 151}, 114 (1999). 


%\bibitem{Valence} J.M. Lawrence, M.C. Croft and R.D. Parkes, 
% Phys. Rev. Lett. {\bf 35}, 289 (1975).  
%\bibitem{Onuki_valence} A. Onuki, Phys. Rev. {\bf B39}, 12308 (1989).

%\bibitem{Gammon} E. Courtens, R. Gammon  and 
%S. Alexander, Phys. Rev. Lett. {\bf 43}, 1026 (1979); 
% E. Courtens and R.W. Gammon, Phys. Rev. B {\bf 24}, 3890 (1981).





\end{thebibliography}
\end{document}